%% file: Predicting_Neutrino_Emission_for_the_Sources_in_the_HESS_Galactic_Plane_Survey.tex
\documentclass[a4paper,11pt]{article}
\usepackage{pos}

\usepackage[utf8]{inputenc}
\usepackage{amsmath}
\usepackage{amsfonts}
\usepackage{amssymb}
\usepackage{graphicx}

\usepackage{txfonts}
\usepackage{url,hyperref}
\usepackage{subfig}

\usepackage{natbib}

\graphicspath{ {./figures/} }

\usepackage{longtable}
\usepackage[figuresleft]{rotating}

\input{includes}

\title{"Predicting Neutrino Emission for the Sources in the H.E.S.S. Galactic Plane Survey"}

\author*[a]{Rowan Batzofin}
\author[a]{Nukri Komin}

\affiliation[a]{University of the Witwatersrand,\\
  1 Jan Smuts ave, Johannesburg, South Africa}
  
\emailAdd{1065203@students.wits.ac.za}
\emailAdd{nukri.komin@wits.ac.za}
\abstract{The H.E.S.S. Galactic Plane Survey has detected very-high-energy (VHE) gamma-ray emission from 78 sources in the Milky Way. These sources belong to different object classes (pulsar wind nebulae, supernova remnants or binary systems) and some of these sources remain unidentified. The gamma-ray emission of these objects may be of leptonic or hadronic origin and gamma-ray observations alone cannot distinguish between these two scenarios. The detection of neutrino emission would provide evidence for a hadronic scenario in these objects.\paragraph*{}
Based on the observed gamma-ray spectra we predict the neutrino emission of these sources under the hypothesis that the emission is solely of hadronic origin. This prediction relies entirely on observation and is independent of the source class, the distance or the ambient target material. We use these predictions to create an empirical model for the neutrino emission of the Milky Way. This model can be used to search for neutrino emission from individual gamma-ray sources as well as testing for neutrino emission from potential source populations in the Milky Way.}

\FullConference{37$^{\rm{th}}$ International Cosmic Ray Conference (ICRC 2021)\\
		July 12th -- 23rd, 2021\\
		Online -- Berlin, Germany}

\begin{document}
\maketitle

%\begin{abstract}
%\noindent
%The H.E.S.S. Galactic Plane Survey has detected very-high-energy (VHE) gamma-ray emission from 78 sources in the Milky Way. These sources belong to different object classes (pulsar wind nebulae, supernova remnants or binary systems) and some of these sources remain unidentified. The gamma-ray emission of these objects may be of leptonic or hadronic origin and gamma-ray observations alone cannot distinguish between these two scenarios. The detection of neutrino emission would provide evidence for a hadronic scenario in these objects.
%\paragraph*{}
%Based on the observed gamma-ray spectra we predict the neutrino emission of these sources under the hypothesis that the emission is solely of hadronic origin. This prediction relies entirely on observation and is independent of the source class, the distance or the ambient target material. We use these predictions to create an empirical model for the neutrino emission of the Milky Way. This model can be used to search for neutrino emission from individual gamma-ray sources as well as testing for neutrino emission from potential source populations in the Milky Way.

%\end{abstract}
\section{Introduction}
The H.E.S.S. Galactic plane survey is a very-high-energy gamma-ray survey in the Galactic plane from $250^{\circ}$ to $64^{\circ}$ galactic longitude and galactic latitude $\leq 3$ \citep{HESS_Galactic_plane_survey}. There are 78 sources in this survey but 48 of them are unidentified. This opens the question as to what is producing these gamma rays. Neutrino emission being detected coming form these sources would be evidence of the gamma-ray emission being of hadronic origin. In this work it is assumed that all the emission (unless already known to be leptonic) is of hadronic origin in order to predict possible neutrino emission.  

\paragraph*{}
Parametrizations for the energy spectra of secondary particles ($\pi$ mesons, gamma rays, electrons and neutrinos) produced in proton-proton collisions are presented in \citep{Kelner_paper}. These parametrizations can be used for energy distributions of secondary particles in the energy range $>$ 100 GeV and have an accuracy better than 10\%. These parametrizations can be used to model the gamma-ray and neutrino emission from proton-proton collisions. This was used to create our model from which the gamma-ray data is fitted and then the prediciton of neutrino emission is made.

\paragraph*{}
Naima is a python package that computes the non-thermal radiation from relativistic particle populations, it also has tools for performing MCMC (Markov Chain Monte Carlo) fitting of radiative models to observed spectra \citep{naima}. The models available are: Synchrotron, Inverse Compton, Bremsstrahlung and neutral pion decay processes.

\paragraph*{}
Naima provides models for gamma-ray emission from proton-proton interactions based on the parametrizations of \citep{Naima_Kafexhiu} and \citep{Kelner_paper}. In this work we make use of the parametrisation of \citep{Kelner_paper} corresponding to our neutrino emission model. %This model is based on analytic parametrizations to the energy spectra and production rates of gamma rays from proton-proton interactions developed in \citep{Naima_Kafexhiu}. This is only able to model the gamma ray emission and not the neutrino emission. The accuracy of this model is better than 20\%. This model can be used for energies $>$ 2 GeV. This model is fitted the H.E.S.S. gamma-ray data.

%\paragraph*{}
%Modelling the gamma-ray spectrum using Naima produces a model better than 20\% while the Kelner model produces a model of better than 10\%. We would need to first use the Naima model \citep{Naima_Kafexhiu} to determine the proton population (giving us a 20\% model) and then use that to predict the neutrino spectrum using the Kelner model \citep{Kelner_paper} (which would give a final model with an accuracy of around 2\%). 

%\section{Models}
%\subsection{Kelner}
%\citep{Kelner_paper} describes analytical formulas for simulated results of the energy spectra of secondary particles produced in proton-proton collisions. These analytical formulas deviate by just a few percent from the simulated distributions over a large range of energies.
%\subsection{Naima}
%Naima is a python package that computes the non-thermal radiation from relativistic particle populations, it also has tools for performing MCMC fitting of radiative models to observed spectra Naima\citep{naima}. The models available are: Synchrotron, Inverse Compton, Bremsstrahlung and neutral pion decay processes. 
%\paragraph*{}
%For modelling hadronic emission, Naima has a Pion Decay model. This model is based on analytic parametrizations to the energy spectra and production rates of gamma rays from proton-proton interactions developed in \citep{Naima_Kafexhiu}. This is only able to model the gamma ray emission and not the neutrino emission. The accuracy of this model is better than 20\%. This model can be used for energies $>$ 2 GeV.

\section{Method}
\subsection{Neutrino flux}
The first step was to make a model for each of the sources in the H.E.S.S. Galactic Plane Survey. For sources that had already been identified as PWN (Pulsar Wind Nebula) no neutrino prediction was made as the emission is supposed to be leptonic. For all other sources the emission was assumed to be hadronic. In total 66 of the 78 H.E.S.S. sources were used. A parametrisation of the proton spectrum for each source is made using naima \citep{naima} and the model of Kelner et al \citep{Kelner_paper} based on the gamma-ray radiation. This proton spectrum is then used to get a prediction of the neutrino flux. If there is a Fermi \citep{4FGL} source nearby to the H.E.S.S. source, the combined H.E.S.S. and Fermi gamma-ray emission is used to make the fit unless the fit was significantly better by excluding the Fermi source and therefore highly unlikely that the Fermi source is a counterpart to the H.E.S.S. source. The fitting done with naima uses an affine-invariant ensemble sampler for Markov Chain Monte Carlo. Three parameters are used to define the proton spectrum: normalisation, spectral index and exponential cutoff energy.

\begin{figure}[h]%
\centering
\includegraphics[width = \textwidth]{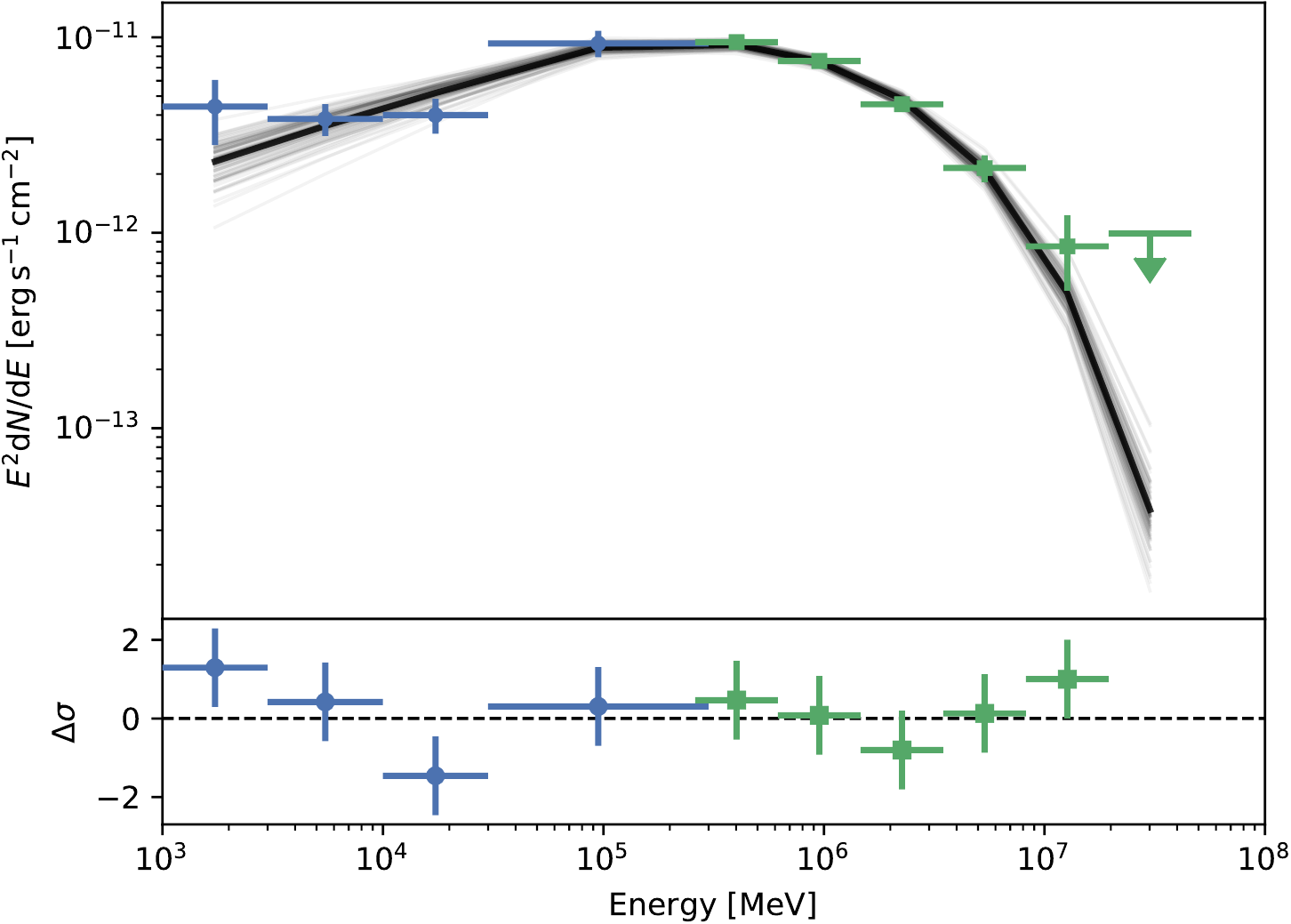}
\caption{A plot of the samplers fit for HESS J1640-465. The blue points are for the Fermi source 4FGL J1640.6-4632 and the green points for HESS J1640-465. The dark black line is for the parameters with the maximum log likelihood value of the samplers. The $\chi^{2}$ p-value is 0.34}
\label{fig:sampler}
\end{figure}

\begin{figure}[h]%
\centering
\includegraphics[width = \textwidth]{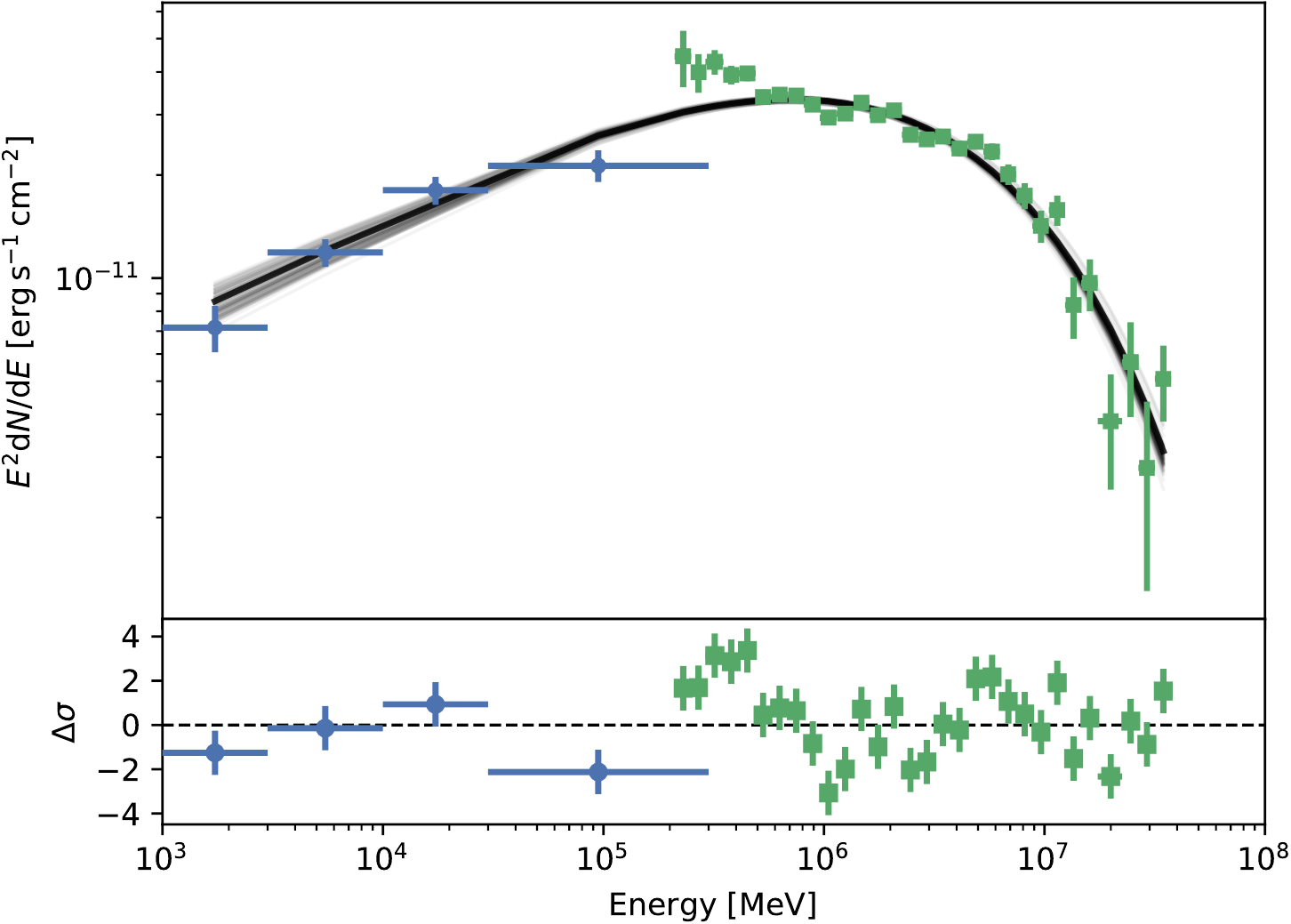}
\caption{A plot of the samplers fit for HESS J1713-397. The blue points are for the Fermi source 4FGL J1713.5-3945e and the green points for HESS J1713-397. The dark black line is for the parameters with the maximum log likelihood value of the samplers. The $\chi^{2}$ p-value is $4.2 \times 10^{-8}$}
\label{fig:worst_sampler}
\end{figure}

\paragraph*{}
Fig.~\ref{fig:sampler} and Fig.~\ref{fig:worst_sampler} shows the gamma-ray emission of HESS~J1640-465 and HESS~J1713-397 respectively and gives examples of what the happens in the fitting. The thick black line shows the line with the maximum log likelihood while the gray lines are the other samples. In this fit all the samples are quite close together and fit the data reasonably well.

\subsection{Creating the map}
To create the neutrino map gammapy,\footnote{https://www.gammapy.org} a community-developed core Python package for TeV gamma-ray astronomy \citep{gammapy:2017, gammapy:2019} is used. A blank map is made with the same dimensions as the H.E.S.S. Galactic Plane map with an added energy dimension, the energy axis ranges from 1~TeV to 100~TeV with 4 bins: 1~TeV to 3.2~TeV, 3.2~TeV to 10~TeV, 10~TeV to 31.6~TeV and 31.6~TeV to 100~TeV. A Sky Model is then created for each source, this has two parts a spatial model and a spectral model. The spectral model is the muon neutrino part of our model based on \citep{Kelner_paper}. The spatial model is created using the information from the H.E.S.S. Galactic Plane Survey results, each source is spatially modelled as either a point source, a shell model, a Gaussian or x-Gaussian (where the source is made up of multiple gaussian components). The map is then populated by taking the integral of the muon neutrino flux over each energy bin.

\section{Discussion} 
Table~\ref{tab:Parameters} lists the best-fit parameters for the proton spectra, the predicted muon neutrino flux and the $\chi^{2}$ p-value of the fit for each source. The maximum neutrino flux over the energy range 1~TeV to 100~TeV is $2.98 \times{ 10^{-11}}$~$\mathrm{TeV\,s^{-1}\,cm^{-2}}$ for HESS~J0852-463. The $\chi^{2}$ p-values are calculated using the loglikelihood values determined during the fitting with naima, the loglikelihood is related to $\chi^{2}$ by $\chi^{2}$ = -2 ln$\mathcal{L}$. The worst $\chi^{2}$ p-value is $4.2\times10^{-8}$ for HESS J1713-397, the mean is 0.342 and the maximum is 0.962 for source HESS J1841-055. Fig.~\ref{fig:worst_sampler} shows the plot of the samplers for HESS J1713-397. It has a low $\chi^{2}$ p-value because very few of the data points lie on the best fit line but instead lie just above or just below. Fitting only HESS J1713-397 does improve the fit but it would still be the worst fit of our sources with a $\chi^{2}$ p-value of $1.3\times10^{-5}$.

\begin{sidewaysfigure}[ht!]
\centering
\includegraphics[scale = 0.4]{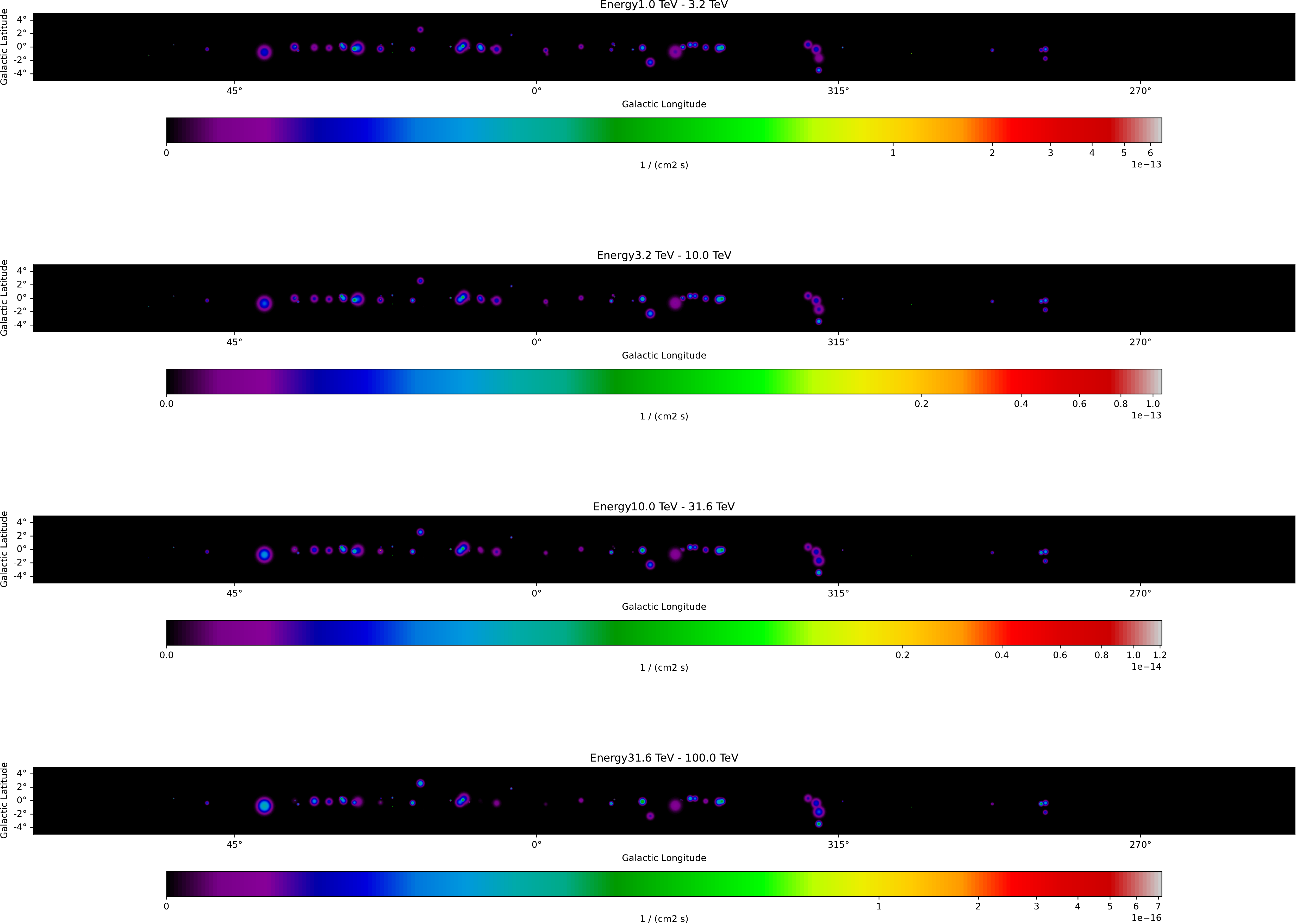}

\caption{Predicted integrated neutrino flux of the Milky Way per energy bin.}
\label{fig:Neutrino_map}
\end{sidewaysfigure}

\paragraph*{}
The predicted neutrino emission map of the Milky Way can be seen in Fig.~\ref{fig:Neutrino_map}. The map is split into the four energy bins: 1 TeV to 3.2 TeV, 3.2 TeV to 10 TeV, 10 TeV to 31.6 TeV and 31.6 TeV to 100 TeV. In total 66 sources are plotted but 9 cannot be seen because their integrated neutrino fluxes are too small to fit into the scale of the map shown here. The map is produced using only observations and does not make use of the source class, distance to the source or the ambient target material.

\clearpage

\section{Conclusion}
We have paremetrised the proton population and created a prediction of the neutrino flux for 66 sources in the H.E.S.S. Galactic Plane survey. Of the 66 sources we fitted 19 included data from a 4FGL\citep{4FGL} source. The neutrino emission at higher energies strongly depends on the cut-off of the proton spectrum. The cut-off is constrained by the gamma-ray spectrum for some but not all of the sources. The neutrino emission of the sources without constrained cut-off is certainly overestimated. This model can be used as a template to be fit to neutrino observational data and to test for neutrino emission from known gamma-ray sources. Neutrino emission from the Milky Way can also arise from interactions of the sea of Cosmic Rays with the interstellar material. Diffuse gamma-ray emission detected by H.E.S.S. \cite{PhysRevD.90.122007} and the Tibet Air Shower Array \cite{PhysRevLett.126.141101} can be used to predict diffuse neutrino emission but is not considered in this work.

\begin{center}
\footnotesize
\begin{longtable}{|c|c|c|c|c|c|}
\caption{Table of fitted parameters for the proton spectra. The Neutrino flux is the Neutrino flux in the energy range 1~TeV to 100~TeV.} \label{tab:Parameters} \\
\hline \multicolumn{1}{|c|}{\textbf{Source Name}} & \multicolumn{1}{c|}{\textbf{normalisation}} & \multicolumn{1}{c|}{\textbf{Spectral}} & \multicolumn{1}{c|}{\textbf{Cutoff}} & \multicolumn{1}{c|}{\textbf{Neutrino flux}} & \multicolumn{1}{c|}{\textbf{$\boldsymbol\chi^{2}$}} \\
\multicolumn{1}{|c|}{\textbf{}} & \multicolumn{1}{c|}{\textbf{$\mathrm{TeV^{-1}\,s^{-1}\,cm^{-2}}$}} & \multicolumn{1}{c|}{\textbf{Index}} & \multicolumn{1}{c|}{\textbf{$\mathrm{TeV}$}} & \multicolumn{1}{c|}{\textbf{$\mathrm{TeV\,s^{-1}\,cm^{-2}}$}} & \multicolumn{1}{c|}{\textbf{p-value}} \\
 \hline
\endfirsthead

\multicolumn{6}{c}%
{{\bfseries \tablename\ \thetable{} -- continued from previous page}} \\
\hline \multicolumn{1}{|c|}{\textbf{Source Name}} & \multicolumn{1}{c|}{\textbf{normalisation}} & \multicolumn{1}{c|}{\textbf{Spectral}} & \multicolumn{1}{c|}{\textbf{Cutoff}} & \multicolumn{1}{c|}{\textbf{Neutrino flux}} & \multicolumn{1}{c|}{\textbf{$\boldsymbol\chi^{2}$}} \\
\multicolumn{1}{|c|}{\textbf{}} & \multicolumn{1}{c|}{\textbf{$\mathrm{TeV^{-1}\,s^{-1}\,cm^{-2}}$}} & \multicolumn{1}{c|}{\textbf{Index}} & \multicolumn{1}{c|}{\textbf{$\mathrm{TeV}$}} & \multicolumn{1}{c|}{\textbf{$\mathrm{TeV\,s^{-1}\,cm^{-2}}$}} & \multicolumn{1}{c|}{\textbf{p-value}} \\
 \hline 
\endhead

\hline \multicolumn{6}{|r|}{{Continued on next page}} \\ \hline
\endfoot

\hline
\endlastfoot

HESS J0852-463 & (2.40$\pm$0.7)$\times 10^{48}$ & 0.897$\pm$0.12 & 21.11$\pm$2.6 & (2.98 $\pm$1.7)$\times 10^{-11}$ & 0.887 \\
HESS J1018-589 A & (5.67$\pm$5.5)$\times 10^{46}$ & 1.623$\pm$0.23 & 118.27$\pm$29.2 & (4.22$\pm$6.3)$\times 10^{-13}$ & 0.276 \\
HESS J1018-589 B & (8.31$\pm$8.3)$\times 10^{46}$ & 1.298$\pm$0.32 & 68.89$\pm$20.3 & (1.37$\pm$2.7)$\times 10^{-12}$ & 0.922 \\
HESS J1023-575 & (2.65$\pm$0.3)$\times 10^{48}$ & 2.083$\pm$0.03 & 156.41$\pm$20.4 & (3.64$\pm$0.6)$\times 10^{-12}$ & 0.023 \\
HESS J1026-582 & (6.64$\pm$8.8)$\times 10^{44}$ & 0.207$\pm$0.28 & 85.92$\pm$47.4 & (2.24$\pm$5.6)$\times 10^{-12}$ & 0.078 \\
HESS J1119-614 & (2.84$\pm$1.1)$\times 10^{48}$ & 2.376$\pm$0.17 & 93.46$\pm$34.5 & (1.02$\pm$0.8)$\times 10^{-12}$ & 0.142 \\
HESS J1302-638 & (2.34$\pm$1.3)$\times 10^{48}$ & 2.740$\pm$0.19 & 473.77$\pm$130.1 & (3.81$\pm$3.6)$\times 10^{-13}$ & 0.003 \\
HESS J1427-608 & (2.64$\pm$55.7)$\times 10^{48}$ & 2.617$\pm$0.97 & 151.29$\pm$60.3 & (5.09$\pm$0.02)$\times 10^{-13}$ & 0.720 \\
HESS J1442-624 & (1.53$\pm$0.2)$\times 10^{48}$ & 1.554$\pm$0.06 & 22.66$\pm$2.4 & (2.40$\pm$0.7)$\times 10^{-12}$ & 0.068 \\
HESS J1457-593 & (1.55$\pm$1.0)$\times 10^{49}$ & 2.519$\pm$0.25 & 305.38$\pm$62.8 & (5.15$\pm$1.1)$\times 10^{-12}$ & 0.232 \\
HESS J1458-608 & (2.47$\pm$55.7)$\times 10^{47}$ & 1.521$\pm$1.01 & 159.98$\pm$90.0 & (3.75$\pm$306.1)$\times 10^{-12}$ & 0.034 \\
HESS J1503-582 & (1.75$\pm$26.2)$\times 10^{49}$ & 2.680$\pm$0.37 & 274.15$\pm$74.5 & (3.19$\pm$59.2)$\times 10^{-12}$ & 0.072 \\
HESS J1507-622 & (1.37$\pm$0.1)$\times 10^{48}$ & 1.876$\pm$0.03 & 218.01$\pm$43.0 & (5.40$\pm$1.1)$\times 10^{-12}$ & 0.905 \\
HESS J1534-571 & (8.94$\pm$4.0)$\times 10^{47}$ & 1.633$\pm$0.22 & 53.57$\pm$21.8 & (2.99$\pm$3.4)$\times 10^{-12}$ & 0.146 \\
HESS J1614-518 & (5.43$\pm$0.4)$\times 10^{48}$ & 1.843$\pm$0.02 & 48.75$\pm$11.8 & (7.78$\pm$1.8)$\times 10^{-12}$ & 0.158 \\
HESS J1616-508 & (7.84$\pm$1.9)$\times 10^{48}$ & 2.030$\pm$0.08 & 126.53$\pm$5.2 & (1.17$\pm$0.5)$\times 10^{-11}$ & 0.037 \\
HESS J1626-490 & (4.60$\pm$36.1)$\times 10^{48}$ & 2.330$\pm$0.46 & 358.46$\pm$122.8 & (3.34$\pm$0.5)$\times 10^{-12}$ & 0.495 \\
HESS J1632-478 & (6.80$\pm$0.3)$\times 10^{48}$ & 2.416$\pm$0.01 & 230.45$\pm$19.2 & (3.06$\pm$0.2)$\times 10^{-12}$ & 0.166 \\
HESS J1634-472 & (4.77$\pm$1.0)$\times 10^{48}$ & 2.277$\pm$0.08 & 477.42$\pm$113.3 & (4.69$\pm$1.8)$\times 10^{-12}$ & 0.043 \\
HESS J1640-465 & (3.89$\pm$0.3)$\times 10^{48}$ & 1.696$\pm$0.04 & 15.69$\pm$1.1 & (2.37$\pm$0.4)$\times 10^{-12}$ & 0.430 \\
HESS J1641-463 & (7.49$\pm$0.6)$\times 10^{47}$ & 2.569$\pm$0.02 & 321.17$\pm$34.3 & (2.1$\pm$0.2)$\times 10^{-13}$ & 0.036 \\
HESS J1646-458 & (2.74$\pm$38.2)$\times 10^{49}$ & 2.643$\pm$0.40 & 239.32$\pm$43.3 & (5.50$\pm$1.0)$\times 10^{-12}$ & 0.368 \\
HESS J1702-420 & (2.35$\pm$10.1)$\times 10^{48}$ & 1.916$\pm$0.5 & 336.01$\pm$89.0 & (9.82$\pm$0.7)$\times 10^{-12}$ & 0.074 \\
HESS J1708-410 & (4.43$\pm$2.2)$\times 10^{47}$ & 1.131$\pm$0.27 & 9.85$\pm$2.7 & (6.12$\pm$6.9))$\times 10^{-13}$ & 0.641 \\
HESS J1708-443 & (3.67$\pm$2.0)$\times 10^{46}$ & 0.301$\pm$0.13 & 27.28$\pm$7.2 & (6.93$\pm$6.6)$\times 10^{-12}$ & 0.129 \\
HESS J1713-381 & (1.61$\pm$0.2)$\times 10^{48}$ & 1.956$\pm$0.0.05 & 14.51$\pm$1.8 & (4.22$\pm$1.0)$\times 10^{-13}$ & 0.407 \\
HESS J1713-397 & (9.49$\pm$0.3)$\times 10^{48}$ & 1.768$\pm$0.02 & 63.74$\pm$3.2 & (2.28$\pm$0.2)$\times 10^{-11}$ & 4.2$\times 10^{-8}$ \\
HESS J1714-385 & (6.37$\pm$2.1)$\times 10^{47}$ & 2.452$\pm$0.11 & 352.44$\pm$91.4 & (2.84$\pm$1.6)$\times 10^{-13}$ & 0.340 \\
HESS J1718-374 & (1.17$\pm$0.2)$\times 10^{48}$ & 2.440$\pm$0.03 & 10.92$\pm$41.0 & (6.26$\pm$11.4)$\times 10^{-14}$ & 0.053 \\
HESS J1718-385 & (1.07$\pm$3.4)$\times 10^{37}$ & -4.596$\pm$0.75 & 20.53$\pm$3.7 & (2.15$\pm$27.7)$\times 10^{-12}$ & 0.281 \\
HESS J1729-345 & (1.88$\pm$34.1)$\times 10^{48}$ & 2.314$\pm$1.71 & 142.51$\pm$53.7 & (1.02$\pm$337.7)$\times 10^{-12}$ & 0.312 \\
HESS J1731-347 & (2.14$\pm$0.5)$\times 10^{48}$ & 1.963$\pm$0.10 & 44.24$\pm$10.2 & (1.86$\pm$0.8)$\times 10^{-12}$ & 0.894 \\
HESS J1741-302 & (2.50$\pm$3.8)$\times 10^{47}$ & 2.250$\pm$0.52 & 440.43$\pm$88.6 & (2.68$\pm$13.6)$\times 10^{-13}$ & 0.581 \\
HESS J1745-290 & (2.54$\pm$0.1)$\times 10^{48}$ & 2.159$\pm$0.03 & 109.91$\pm$5.7 & (2.14$\pm$0.2)$\times 10^{-12}$ & 0.376 \\
HESS J1745-303 & (3.87$\pm$0.2)$\times 10^{48}$ & 2.371$\pm$0.01 & 56.57$\pm$7.6 & (1.06$\pm$0.1)$\times 10^{-12}$ & 0.025 \\
HESS J1746-285 & (4.12$\pm$12.2)$\times 10^{46}$ & 1.702$\pm$0.76 & 287.30$\pm$68.0 & (4.26$\pm$60.6)$\times 10^{-13}$ & 0.312 \\
HESS J1746-308 & (7.89$\pm$2.0)$\times 10^{48}$ & 3.373$\pm$0.33 & 84.35$\pm$24.6 & (1.42$\pm$1.4)$\times 10^{-13}$ & 0.413 \\
HESS J1747-248 & (4.15$\pm$83.8)$\times 10^{47}$ & 2.232$\pm$0.61 & 339.88$\pm$82.5 & (4.42$\pm$142.0)$\times 10^{-13}$ & 0.164 \\
HESS J1800-240 & (8.63$\pm$70.2)$\times 10^{48}$ & 2.455$\pm$0.21 & 365.66$\pm$130.1 & (3.84$\pm$30.9)$\times 10^{-12}$ & 0.639 \\
HESS J1801-233 & (3.04$\pm$0.2)$\times 10^{48}$ & 2.643$\pm$0.02 & 30.20$\pm$18.2 & (2.48$\pm$1.0)$\times 10^{-13}$ & 0.428 \\
HESS J1804-216 & (2.37$\pm$0.5)$\times 10^{49}$ & 2.342$\pm$0.16 & 25.92$\pm$60.7 & (3.97$\pm$4.7)$\times 10^{-12}$ & 0.014 \\
HESS J1808-204 & (1.73$\pm$4.8)$\times 10^{47}$ & 2.091$\pm$0.57 & 410.02$\pm$141.2 & (3.61$\pm$24.7))$\times 10^{-13}$ & 0.028 \\
HESS J1809-193 & (7.40$\pm$2.0)$\times 10^{48}$ & 2.107$\pm$0.10 & 128.01$\pm$35.2 & (8.27$\pm$3.9)$\times 10^{-12}$ & 0.186 \\
HESS J1813-126 & (2.91$\pm$0.6)$\times 10^{47}$ & 1.737$\pm$0.49 & 318.53$\pm$79.9 & (2.72$\pm$12.9)$\times 10^{-12}$ & 0.941 \\
HESS J1813-178 & (2.23$\pm$0.7)$\times 10^{47}$ & 1.178$\pm$0.11 & 45.58$\pm$18.3 & (3.44$\pm$2.7)$\times 10^{-12}$ & 0.141 \\
HESS J1826-130 & (2.49$\pm$1.8)$\times 10^{47}$ & 1.579$\pm$0.22 & 155.57$\pm$63.1 & (2.85$\pm$3.9)$\times 10^{-12}$ & 0.198 \\
HESS J1826-148 & (9.55$\pm$3.1)$\times 10^{47}$ & 2.066$\pm$0.15 & 146.50$\pm$49.5 & (1.35$\pm$1.0)$\times 10^{-12}$ & 0.398 \\
HESS J1828-099 & (3.20$\pm$55.0)$\times 10^{47}$ & 2.018$\pm$0.70 & 201.38$\pm$65.8 & (6.58$\pm$268.1)$\times 10^{-13}$ & 0.304 \\
HESS J1832-085 & (2.33$\pm$1.8)$\times 10^{47}$ & 2.017$\pm$0.23 & 163.30$\pm$35.1 & (4.27$\pm$5.6)$\times 10^{-13}$ & 0.522 \\
HESS J1832-093 & (2.88$\pm$21.7)$\times 10^{47}$ & 2.028$\pm$0.67 & 36.97$\pm$124.4 & (1.71$\pm$40.2)$\times 10^{-13}$ & 0.311 \\
HESS J1833-105 & (4.55$\pm$80.6)$\times 10^{47}$ & 2.175$\pm$1.20 & 132.97$\pm$93.4 & (4.01$\pm$376.3)$\times 10^{-13}$ & 0.246 \\
HESS J1834-087 & (5.59$\pm$0.6)$\times 10^{48}$ & 2.145$\pm$0.02 & 46.25$\pm$2.7 & (2.76$\pm$0.2)$\times 10^{-12}$ & 0.089 \\
HESS J1841-055 & (1.86$\pm$0.4$\times 10^{49}$ & 2.104$\pm$0.14 & 60.05$\pm$46.8 & (1.28$\pm$1.0)$\times 10^{-11}$ & 0.962 \\
HESS J1843-033 & (7.33$\pm$6.5)$\times 10^{47}$ & 1.520$\pm$0.27 & 82.52$\pm$79.8 & (5.94$\pm$11.5)$\times 10^{-12}$ & 0.279 \\
HESS J1844-030 & (4.42$\pm$2.2)$\times 10^{47}$ & 2.079$\pm$0.19 & 77.48$\pm$29.8 & (4.00$\pm$3.6)$\times 10^{-13}$ & 0.886 \\
HESS J1846-029 & (9.84$\pm$172.8)$\times 10^{47}$ & 2.271$\pm$0.56 & 167.89$\pm$58.8 & (6.76$\pm$197.3)$\times 10^{-13}$ & 0.340 \\
HESS J1848-018 & (3.07$\pm$3.7)$\times 10^{48}$ & 2.384$\pm$0.34 & 442.62$\pm$155.8 & (1.89$\pm$4.1)$\times 10^{-12}$ & 0.751 \\
HESS J1852-000 & (6.55$\pm$133.7)$\times 10^{47}$ & 1.925$\pm$0.92 & 316.65$\pm$62.8 & (2.55$\pm$159.7)$\times 10^{-12}$ & 0.243 \\
HESS J1857+026 & (1.13$\pm$0.3)$\times 10^{49}$ & 2.215$\pm$0.11 & 37.12$\pm$7.3 & (3.74$\pm$1.7)$\times 10^{-12}$ & 0.719 \\
HESS J1858+020 & (6.63$\pm$1.0)$\times 10^{47}$ & 2.152$\pm$0.06 & 177.90$\pm$67.9 & (7.41$\pm$2.6)$\times 10^{-13}$ & 0.937 \\
HESS J1908+063 & (6.36$\pm$73.3)$\times 10^{48}$ & 2.012$\pm$0.39 & 303.11$\pm$24.7 & (1.65$\pm$26.6)$\times 10^{-11}$ & 0.343 \\
HESS J1911+090 & (2.19$\pm$22.9)$\times 10^{48}$ & 2.925$\pm$0.64 & 34.07$\pm$123.2 & (9.03$\pm$198.0)$\times 10^{-14}$ & 0.467 \\
HESS J1912+101 & (1.49$\pm$0.3)$\times 10^{48}$ & 1.704$\pm$0.06 & 43.73$\pm$9.1 & (3.11$\pm$1.2)$\times 10^{-12}$ & 0.015 \\
HESS J1923+141 & (1.94$\pm$1.8)$\times 10^{48}$ & 2.426$\pm$0.34 & 281.84$\pm$53.9 & (8.95$\pm$18.2)$\times 10^{-13}$ & 0.428 \\
HESS J1930+188 & (1.00$\pm$7.1)$\times 10^{48}$ & 2.457$\pm$1.13 & 253.33$\pm$153.6 & (4.00$\pm$136.5)$\times 10^{-13}$ & 0.050 \\
HESS J1943+213 & (1.03$\pm$0.2)$\times 10^{48}$ & 1.853$\pm$0.07 & 13.41$\pm$3.5 & (3.24$\pm$1.5)$\times 10^{-13}$ & 0.459 \\
\end{longtable}
\end{center}

%\vspace{-10mm}
\bibliography{Predicting_Neutrino_Emission_for_the_Sources_in_the_HESS_Galactic_Plane_Survey.bib}

\end{document}

%% file: includes.tex
\def\apjl{ApJ}%
\def\apjs{ApJS}%
\def\ao{Appl.~Opt.}%
\def\aap{A\&A}%
\def\utw{\smash{\rlap{\lower5pt\hbox{$\sim$}}}}
\def\udtw{\smash{\rlap{\lower6pt\hbox{$\approx$}}}}

\def\sun{{\lower-2pt\hbox{$_\odot$}}}

\def\g{$\gamma$}